# A Time Localization System in Smart Home Using Hierarchical Structure and Dynamic Frequency

Qichuan Yang
School of Computer Science and Engineering
Beihang University
Beijing, China
yangqichuan@buaa.edu.cn

Zhiqiang He
Lenovo Group
Beijing, China
hezq@lenovo.com

Kai Zhao
Cloud and Intelligent Computing Lab
Lenovo (Beijing) Ltd.
Beijing, China
zhaokai6@lenovo.com

Tian Gao
Lenovo (Beijing) Ltd.
Beijing, China
gaotian1@lenovo.com

*Abstract*—Both GPS and WiFi based localization have been exploited in recent years, yet most researches focus on localizing at home without environment context. Besides, the near home or workplace area is complex and has little attention in smart home or IOT. Therefore, after exploring the realistic route in and out of building, we conducted a time localization system (TLS) based on off-the-shelf smart phones with WiFi identification. TLS can identify the received signal strength indication (RSSI) of home and construct radio map of users' time route without site survey. As designed to service the smart devices in home, TLS applies the time interval as the distance of positions and as the variables of WiFi environment to mark time points. Experimental results with real users show that TLS as a service system for timeline localization achieves a median accuracy of 70 seconds and is more robust compared with nearest neighbor localization approach.

*Keywords-smart home; time localization; RSSI; mobile computing*

## I. INTRODUCTION

Smartphones are rapidly equipped with more sensors since iPhone's birth which makes the location based service (LBS) an important role in digital worlds. LBS has been studied over the past years with the rise of mobile computing. A lot of researches have been attracted to mine the location and route records. Numerous algorithms are introduced from traditional mining methods to this region. They are aimed to advertise the user, to provide interesting attraction or to improve the apps' service.

The existing localization application generally focus on two scenes: outdoor GPS and indoor WiFi. For example, Google Map and Baidu Map can navigate and locate users by GPS, cell tower and connected WiFi. The Apple's indoor localization app, Indoor Survey, is able to determine a user's position by RF signals from access points (AP) [1]. Many IOT devices are able to perform personalized service based on the basic location services.

However, there are still many particular issues need to be studied. For example, GPS is not sensitive in buildings. Free open WiFi is regional and has potential risk. Sensors in smartphone will consume the battery soon if they sense the environment frequently. Most users deny the 24 hours' data collection considering their privacy [2]. What is more, most researches focus on how to serve the user, and to the best of our knowledge there is no specific system designed mainly for smart home device to localize time.

In order to meet the emerging challenges, this article proposes a time localization system (TLS) to integrate discrete data and to provide services for smart devices in home. The principle of TLS is that the commuting time of a user from any AP to his home is relatively stable with the same transportation. In TLS, the time localization is defined by the difference between instant time and arriving home time. And the basic service set identifier (BSSID) of a user's home WiFi can be mined from daily sampling. Then the arriving home time point can be defined when the BSSID is detected. In order to integrate daily data, a hierarchical structure is applied to response the query of BSSID. By adding new distinct BSSID from daily data continually, a more complete AP's environment of the user is constructed considering different transportation. This hierarchical structure can also keep the BSSID unique and provide a quick response. Since the latest records of the user continuously update the structured data, the TLS could archive a convictive result. More importantly, TLS utilizes a dynamic frequency mechanism to control the sensors and to reduce the calculation in smart phone.

Compared with the existing localization system, our main contributions are as follows:

- We propose a concise time localization system to provide arriving time prediction service for both smart home devices and users. This service can act as a commander of IOT or a profile to define a user.
- We developed a designed dynamic sensing frequency in application (app) to collect the date from on the shelf smart phone sensors in order to save battery. Users are able to use their phones normally without any operation or instruction.
- We design a hierarchical structure to integrate the users' daily records. The first level tags AP with timestamp mined from history records and responses the BSSID query. The second level extracts user's schedule from his daily life.

The rest of the paper is organized as follows. Section II describes the intuition and motivation. Section III provides the related work. Section IV describes the algorithm and system in detail. Section V demonstrates the experimental results followed by the conclusion in section VI.

## II. INTUITION AND MOTIVATION

It may be instructive to consider the real situation of smart home in both the devices' and the users' view firstly. Assume that devices have basic intelligence to manipulate themselves. For the sake of servicing the masters in time they have to schedule their actions according to the masters' arriving time (like a home service in owner's absent). Although installing a RFID in the front door is effective, it is clumsy for some devices that need long time to prepare their works, like boil water or warm the room. Therefor a hierarchy time prediction is necessary for all smart devices.

As to the user, we would not like our moving paths of daily life are recorded and analyzed precisely considering privacy. It is contradictive to the main purpose of prediction. Currently the mainstream of predicting the arriving time is to count the distance from the instant position to home and then to estimate the time with moving speed. Besides the deviation of localization, speed estimation depends on a number of uncertain factors in mobile devices. Thence a safe and precise method to describe the users' routine time is needed.

Although indoor localization of WiFi or Zigbee have achieved inspiring accuracy, the time prediction on users' way to home is still obscure. In the same time smart devices at home are eager to have a relative accurate timeline of their masters to manipulate their daily work rather than to have a precision location. In this aspect, a time localization system is necessary in the smart home environment.

## III. RELATED WORK

Since human's moving pattern have been extensively studied in many areas for years, we seek for the localization system to build a time localization system at first.

To date, mobile computing has attracted a lot of attention. In mobile localization systems, sensors like accelerometer [3], gyroscope [4], magnetometer [4, 5], etc. are used to work with WiFi or GPS [6-8] to improve the accuracy of localization or navigation. In indoor mobile localization researches, information of signal is playing an important role such as RSSI [9, 10], Fingerprint [3, 11], Angle of arrival (AoA) [12, 13], Channel state information (CSI) [14], etc. Besides organic architecture [15] was proposed to discount erroneous WiFi data. Even the backscatter of WiFi [16] was applied as a communication method. Some of these systems require special equipment, multiple antennas or additional knowledge, anchor points determined by human. And the energy consuming is not fully considered for most of this implementation platform is a laptop.

In outdoor localization systems, GPS has been studied to navigate and localize precisely for many years. MIT [17] establishes a project to mine position. They use an app to collect all sensors data from volunteers' Nokia smart phones. Almost 350 thousand hours data are recorded and over one hundred people took part in the project. Microsoft Research

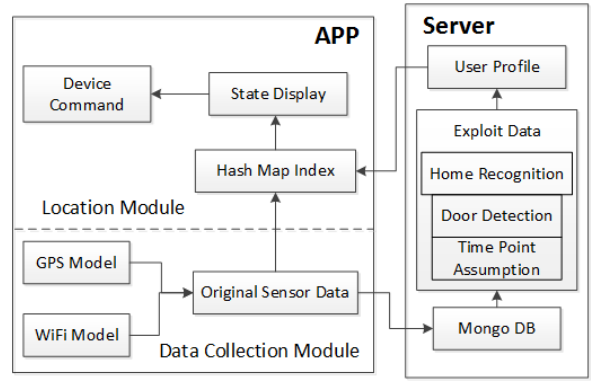

Figure 1. Architecture of TLS

Asia [18-22] also set up a project called Geo-Life to collect GPS data since 2007. They found methods to modify the GPS to increase the accuracy of locating. As the accuracy of civil GPS is limited, they improve location accuracy by the before and after position points [19]. Other people present the moving style and life custom by frequent pattern mining from the GPS [20].

However, as mentioned before, the transition from GPS covering region (outdoor) to rooms in buildings (indoor) is still not smooth. Some researchers are working on it. Yuanchao Shu et al. [23] proposed a system called FOLLOWME to solve the last mile navigation from the gate to the door. Souvik Sen et al. [24] introduced the physical layer information to extract the signal strength and the angle of the direct path to locate users in public place. The proposed TLS inherits using WiFi due to the increasing deployment tendency of WiFi. And TLS skips the physical position but predicts the arriving time directly to support the smart device at home. It can smoothly connect the outdoor and indoor situation.

## IV. SYSTEM OVERVIEW

This section provides an overview of TLS with the definition of time localization at first. Then the section A presents the architecture and the section B illustrates the process stream with a real example.

Definition1: Time Localization (TL) --- the time that user will have to spend to arrive home from the current position.

### A. System Architecture

TLS consists of two parts, the smart phone app and the online server. The app has data collection module and time location module. The server is responsible for the data storage and data process. Figure1 shows the overall architecture of TLS and illustrates the specific architecture of app.

In the app, the location module responses the query of time localization with the user profile and current sensing data of mobile phone. Data collection module records the sensing data and sends them to server for mining. The sensing data contains latitude, longitude, WiFi connection state and scanned environment with timestamp. Note that the collected data is identified by MAC address of user's phone without personal information like name or phone number, and the hierarchical sensing frequency of TLS is displayed in Figure 2-3.

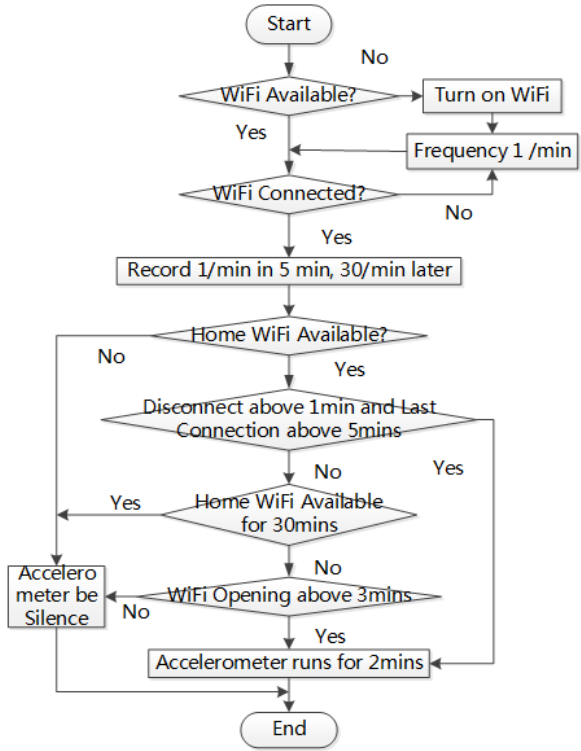

Figure 2. Flowchart of data collection module

The server in Figure 1 accepts original data from smart phone all the time but extract the user profile regularly. User profile is constructed from original data and marked by three parts: BSSID of user's home AP (WiFi), gradient of APs' signal and AP's environment on the way to home. The specific analysis algorithm of original data is shown in V. The result of analysis, user profile, is the source of location module in app. Through mapping user profile to a hierarchical structure, the app can response localization demand and the server could update the profile when the network is available.

For the purpose of battery saving, we propose a dynamic sensing frequency for app. The data flow with frequency is shown in Figure 2. These frequencies vary aiming at detection of reconnection and excluding of useless sensing time if users stay home for a long time. After excluding the abnormal situation, the accelerometer will only be activated when users are in front of their doors. This regulations and parameters are summarized from our original data collection app with certain frequency. There is no endpoint of the flow due to the collection action will run automatically. Comparing with our original app without dynamic frequency, the smart phone can survive about 50% more time in normal life. The frequency also considered the sensors' limited sensing ability in smart phone. For example, the records of WiFi sensing will be constant within 1 second even though the actual WiFi environment is changed.

In the localization module, we design several activation sentences for the app with variable frequency as shown in Figure 3. The GPS region is defined as a 500-meter radius circle area with user's home-centered if GPS could sense position. The reference, 500 meters, is the trigger of the localization. The WiFi region refers to the environment from the edge of GPS's circle to the user's door. Only in the WiFi region the app will start up to localize the arriving time for smart devices.

### B. A Location Example

Suppose a user is going home after work as usual. The app will detect the smart phone every minute to check whether the GPS or WiFi is opened. If sensors are available, the sensing data would be collected. When the user approaches region A shown in Figure 4, the gate of his community, he will step in the rim of GPS region. Then the app's detection frequency of AP is up to every five seconds as shown in Figure 3. All APs' BSSID around the user will be inquired by the location model of Figure 1. This query's result, prediction time 340s, will be sent to devices and shown on user's screen. When the user gets to section C the environment changes to the staircase that contains a vertical distance. The app operates well in the same way as the BSSID environment is diversify in different flow. But GPS always lose signal and cannot distinguish the flows. When the user arrives his floor or near the D region, the BSSID of his home WiFi will be recognized. Accelerometer begins to work and records the user's action for 2 minutes. After the network is stable, the app enters connected state as shown in Figure 2. It would record the data for 5 minutes, then go to sleep and wake up every 30 minutes.

In addition, the update procedure will be started in the midnight if the network is available. The procedure consists of uploading data to serve, removing the outdated data in smart phone and updating the app's user profile. The data in the server is processed daily. The algorithm of this process is illustrated in Section V.A.

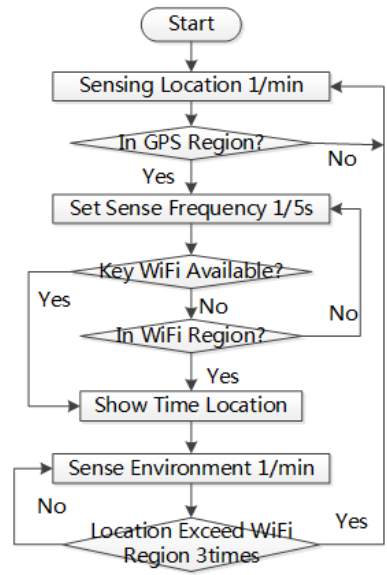

Figure 3. Flowchart of localization module

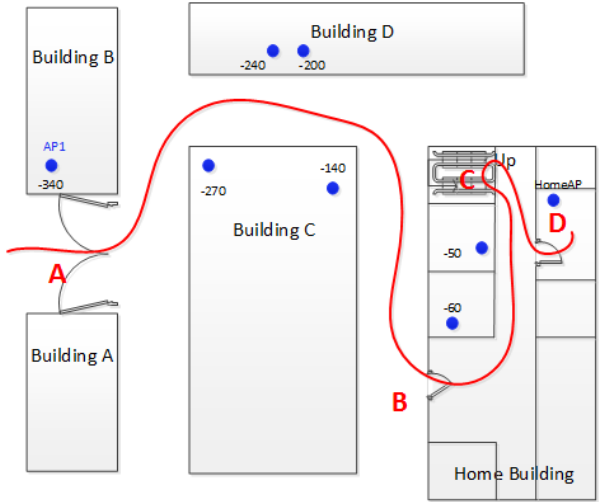

Figure 4. A localization example

*C. Assumptions and Limitatiaons*

As demonstrated above, the TLS can only predict the arriving time if the app detects any BSSID recorded in database. That means the app cannot work well in a new APs' environment. A minimum sliding window is needed to append new environment's data. We use one week as window length in experiment as the weekday is the nature cycle for most people in city. Considering to reduce the calculation redundancy, the fresh data will be calculated solely. The sliding window appends user's data daily.

The smart devices at home are sensitive to time delay. If the devices are not ready when users arrives home, users may be annoyed. But if the device gets instruction a little earlier users may not sense the different. Therefore we extract the nearest time distance as a label to attach each AP.

The existing location methods can also be applied to estimate the arriving time. In this way, the app has to estimate the length of user's step or moving speed, in order to translate the distance to time scale. But the range of moving speed is wide even only considering one transport way in complex traffic of city. That will result in accuracy of time prediction decreases with the increasing distance from home.

## V. ALGORITHM DESIGN

TLS mainly uses three common sensors of smart phone, GPS, WiFi and Accelerometer. WiFi undertakes the scanning function of APs for TLS. Without WiFi the TLS will degenerate to be a dull location system of GPS. The GPS helps to roughly locate the user aiming at supplying a relative coordinate for WiFi. Accelerometer contributes to calculate the specific arriving time point for smart devices. This section illustrates the data processing algorithms running in server with a precondition that all the sensing data needed is accessible and the sliding window is set to seven days.

*A. Structure and Update*

First of all, we need a data structure to duel with the high frequency query and to merge the user's time points. The data in the structure needs to be updated frequently and uniquely. Therefore we introduce a hierarchical structure, two level hash map, as a data container.

*1) Hierarchical Structure*

Hash map has a stable query speed and a unique key list. The key of the hash map is the BSSID of each AP scanned from user's way to home. And the time distance (two labels) between each AP and user's home AP is the value of hash map. By this way, each day has a hash map composed by a series of key-value pairs and the BSSIDs are kept unique naturally.

In order to calculate the time distance, two labels are attached to each AP firstly, the time point that phone loses the AP (Time Point of Losing, TPL) and the duration that the phone can reach the AP (Time Duration of Reaching, TDR). Then we could get the time localization of each AP to subtracting the TPL of home AP by the TPL of each other APs according Definition 1. We now get several pairs of key-value to build the first level hash map for each day's data. The hash map of second level is then can be built with these daily hash maps according to the TDRs of them. The key of the higher level is the difference of their daily TDR taking the first day as standard. In fact this TDRs represent the transportation of users. Algorithm 1 shows the detail of this procedure.

The above process will run in the server every day and the two level structure with data will be duplicated and sent to app when it is online. The latest hash map will response the time localization query. When the app gets TDR of one AP, the higher level hash map will be inquired firstly to find the most similar day (also most similar transportation), then to inquire the second level with the BSSID of AP.

*2) Update mechanism*

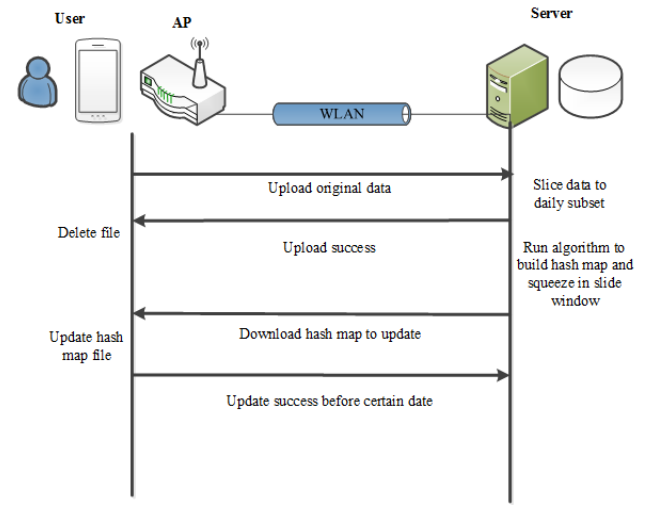

Figure 5. Update mechanism

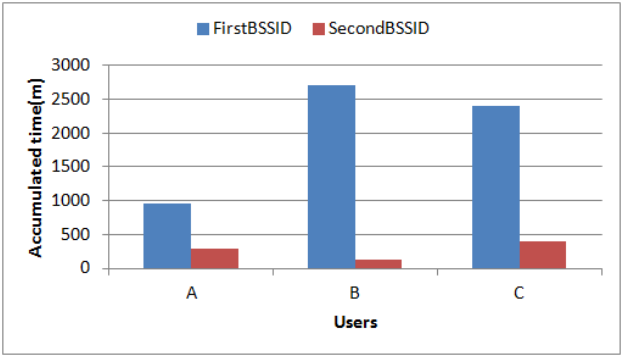

Figure 6. Accumulated time of users' connected BSSID during the nights

In spite of the storage of mobile phone increasing continually, the data that can be stored and used in an app is still limited. Therefor TLS downloads and updates the data structure to app daily when the WLAN network is accessible. Besides, a 7 day's sliding window is used to limit the data exchanged. In this way if a user moves to a new house, the sliding window ensures that TLS will detect it and be updated with new home AP in the 4th day. This mechanism assumes that what will happen in this week is more similar with the last week than earlier. The data before last week will be abstracted as a second choice hash map to response query if the 7 day's two level hash map cannot match any AP's BSSID.

As shown in Figure 5, the server inserts the data to BD and split out the new day's data to run the Algorithm 1 after uploading. The result of algorithm, a hash map with the new day, is sent back to the app. Then file of original data will be removed from phone after downloading. The database of server stores not only the original data of sensors but also the outdated structured data as backup.

*B. Determine Home WiFi*

TLS uses a simple voting method to determine the user's home according to history records. Records is sliced into days firstly. Then every day's TDR of each AP during 21 pm to 6 am written as $w_i(t)$ shown in Equation 1 is compared with each other. After ranging all records, the maximum is assumed as the home AP of that day. Each day votes its home AP and the AP getting the most votes is set as the home AP. Finally the BSSID of this AP will be marked and used in the TLS.

$$AP = \max\{w_1(t), w_2(t) \cdots w_i(t) | t \in [21:00, 06:00]\} \quad (1)$$

Figure 6 shows the accumulated time of three users as example. All the first BSSIDs of these users are significantly ahead the second one and actually the home AP. Hence this method is effective and concise.

*C. Open Door Detection*

Typically, the home AP is connected before the user standing in front of the door. However we found that the actual situation is more complicated. A lot of user's smart phones do not connect to home AP until the user gets in home for a while. Because of the delay when a WiFi model connects to an accessible AP, we define the opening action by AP environment assisted by accelerometer. The scanning operation is manipulated by app as shown in Figure 3.

There are three joint conditions to identify an opening door action:

*1) The home AP is detected, but connection is not necessary;*

*2) Most RSSI of APs including home AP are fluctuating, the accelerometer detectes a standing positon;*

*3) The number of availiable AP is increasing or decreasing;*

As mentioned in [25, 26], the RSSI will fluctuate when the smart phone closes to a corner or a blocking obstacle. We find that this situation also happens when a user crosses the door. The rectangles in Figure 7 (b) are labeled with user's opening door action. It indicates that the RSSI of a home AP has a transition when a user is near the door and opening it. However this feature is not manifest during the entire changing process of RSSI. Therefore TLS also uses the number of scanned APs to determine the time of opening door as shown in Figure 7 (a). The rectangle pointing out the peak of the number series is the actual door opening motion. With the accelerometer's help which can clarify the user's standing action with almost a hundred percent accuracy, the opening action can be detected effectively.

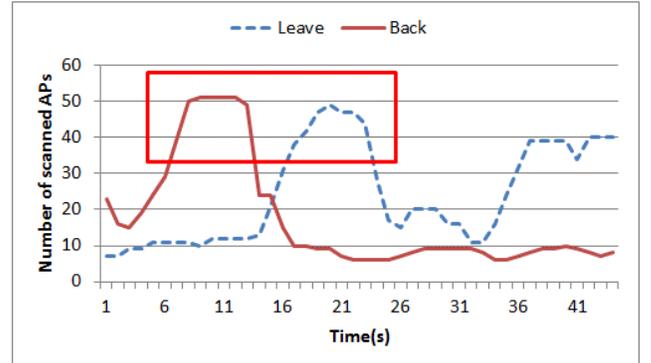

(a) Number of scanned AP wave form for door motion

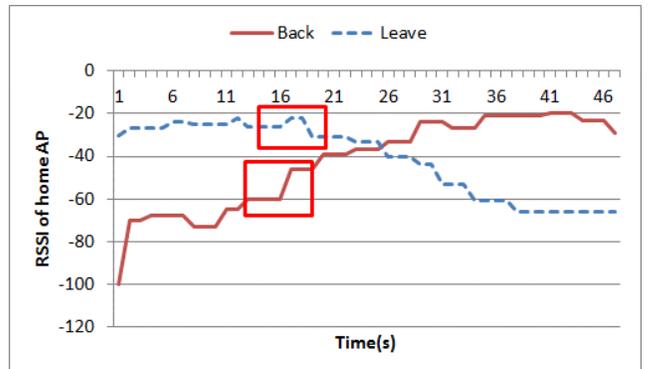

(b) RSSI wave form for door motion

Figure 7. Signal State Fluctuation of Door Motion

**Input:** original records of a day O, user home BSSID $AP_{home}$
**Output:** two level hash map Day Map<Day, AP Map<BSSID, Time Distance>>
**for** each day $D_i$ in O
    **for** each record $R_j$ in $D_i$
        **for** each $AP_m$ in $R_j$
        Calculate time distance between $AP_m$ and $AP_{home}$:
        $tpl_m = (time_{lost\ AP_m} - time_{detect\ AP_{home}})$;
        Calculate duration of $AP_m$:
        $tdr_m = (time_{lost\ AP_m} - time_{detect\ AP_m})$;
        Label $AP_m$;
        Put <$AP_m, (tpl_m, tdr_m)$> in hash map $HM_{D_i}$;
        **end**
    **end**
    Put <$D_i, HM_{D_i}$> in hash map $HM_{lastWeek}$;
**end**

Algorithm 1. Construct function of hash map

The door operation detection will service the smart devices in house like light, TV and air conditioner etc. without additional RFID or Bluetooth. It could active devices at home or let them to be standby.

## VI. EVALUATION

### A. Implementation and Preprocess

We distribute our TLS app to volunteers to install on their phones. These volunteers consist of company employees, university students and temporal interns. Their off-the-shelf smart phones are produced by different companies including Lenovo, Samsung, HTC, Meizu and Xiaomi. They used the app without any instruction. In this case, all data were collected from users' real normal life. The app were kept as a background process if not be stopped manually. Note that the users' smart phones do not have the same Android version, but they are all above 4.4 KitKat.

The APs detected by our app are naturally existing in users' way to home. We do not set up any AP or pick any special firmware of AP for test.

Algorithms written by python runs on online server Linux CentOS 6.5. As offline algorithms, they could be started up any time during a day. There are more than 15 thousands of records in Mongo BD. We choose data of Lenovo K3, Samsung A5 and Samsun Note3 etc. eight users in three months as samples to illustrate TLS's performance after analyzing the corresponding of users' GPS and AP records. We select certain period of them in order to divide the environment into simply and mixture transportation situation.

Since the Home AP detection and the door detection are the base of TLS and already displayed in Section IV, which could provide meaningful and stable information for TLS. In the evaluation we illustrate the performance of TLS, time localization accuracy for service smart devices.

### B. Evaluation

We use the users' data to test how well our system would perform when deployed in simply and complex situations. As TLS is designed to service the smart device at home, we calculate the cumulative distribution function (CDF) by comparing the prediction time with actual time in database. TLS do not predict in the first week of each users.

Specially, the nearest neighbor (NN) is employed as a benchmark system to confirm the effectiveness and improvement. As the distance of NN is defined by AP's environment of each time point, the NN algorithm actually map APs to locate time of route. When a new AP's environment is acquired, NN compares it with all the history time points to get the most similar one. Then the TL of the most similar one is the prediction of that point. Both the sliding window and data of NN and TLS are the same.

Considering AP's environment can be defined by different level of RSSI, we filter the APs in different level to gain a better performance. We compare the performance of different levels of RSSI and plot them in Figure 8. We can see that when we take all accessible APs as AP's environment feature, the error is not decreased while the comparison times are maximum. Therefore we select the -70 dB as the stander of filter for NN and TLS, the bold line in Figure 8.

*1) Simple transportation*

Firstly, we evaluate the algorithms in same users' data. Their transportations and routes are certain according to their GPS records. These users go home straightly after work and this situation was repeated every workday. We strips out these data of simple transportation periods to test. Figure 9 (a) plots the CDF of time localization errors of the results. We can see that the TLS reaches a comparable accuracy compared with the NN. Note that the NN iterates all the records while the TLS only hash two times. The absolute error is more consistent between them as shown in Figure 9 (b). The NN even has a bit of advantage of TLS in the simple transportation because NN

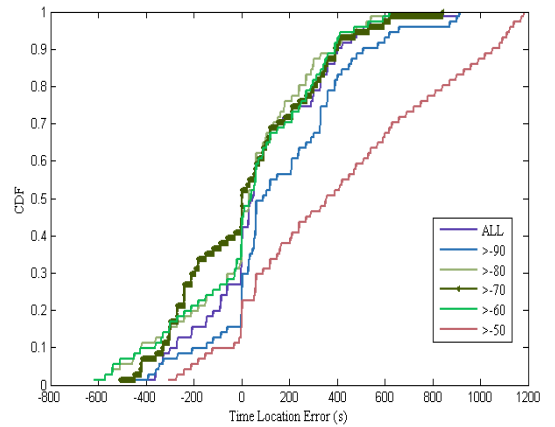

Figure 8. Performance of different RSSI standards

always can obtain the most similar record in history as it iterates all records.

*2) Mixture transportation*

Then we test the performance of TLS in mixture transportation environment. These users randomly walked or cycled to home. Sometimes they even had a dinner outside before arriving home. This situation is more complex and more real than the simple transportation. We handle these mixture data to TLS and NN. As shown in Figure 10, 74% of the TLS's errors are owing to they are ahead the actual time point. The 82% of TLS's errors are in the range of 100 seconds. While the NN trends to delay the time point. The maximum of NN errors is above 200 seconds. The TLS has a better performance than NN in the mixture transportation. Because the NN has more than one most similar records in history but with different timestamps. And the NN cannot distinguish them and has to randomly choose one to output. TLS's advantage is result from the records are classified in days and APs' position are always stable. Besides, a smart device at home would prefer TLS to NN since TLS would give them more seconds to prepare.

## VII. CONCLUSION

In this paper, we present the TLS, a plug-and-play, lightweight prediction system. It provides the time service to the smart device and the user by localizing APs in time line. In particular, it has less calculation pressure for smart phone due to its mapping structure compared with NN algorithm. To the best of our knowledge, TLS is the first system to bridge the service gap between human position and smart devices in time view. TLS will play an important role in the service of smart home with specific functions or more sensors in the future.

## ACKNOWLEDGMENT

This research is supported by CIC research department of Lenovo Ltd. Part of the original data and the applications used in this paper are available on GitHub: www.github.com/ KingAndQueen/Software- Prediction.git.

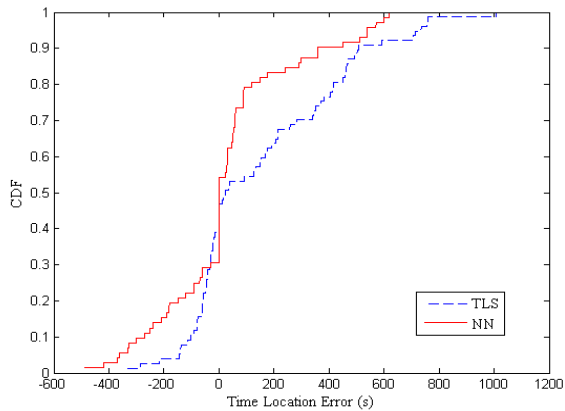

(a) Performance of simple transportation

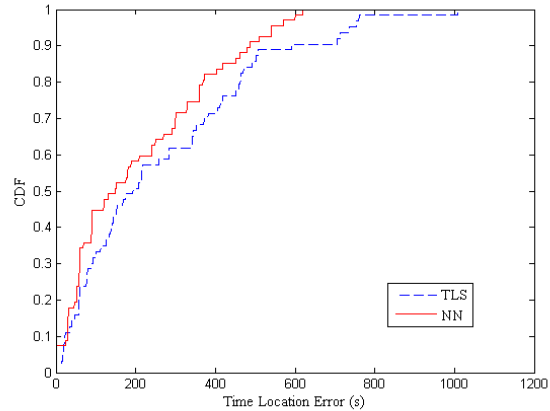

(b) Absolute error of simple transportation

Figure 9. CDF of TLS's localization error compared with NN in the same simple transportation data

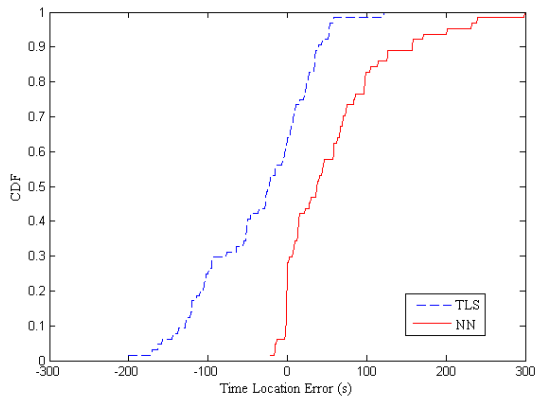 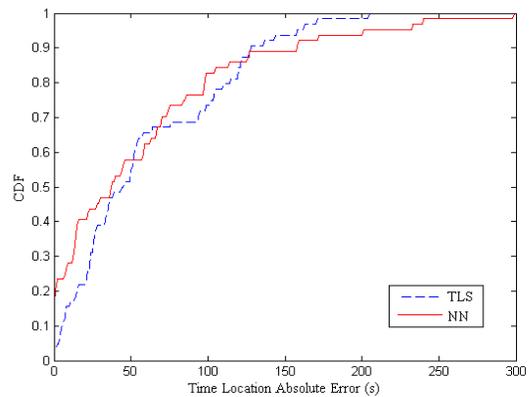

(a) Performance of mixture situation      (b) Absolute error of mixture situation

Figure 10. CDF of TLS's localization error compared with the NN in the same mixture transportation data